\documentclass[final,times,3p,twocolumn]{elsarticle}

\usepackage{lineno,hyperref}
\usepackage{upgreek}
\usepackage{xcolor}
\usepackage{lipsum}  
\usepackage{orcidlink}
\usepackage{soul}
\usepackage{bigdelim,bigdelim, makecell, booktabs}
\modulolinenumbers[1]
\biboptions{numbers,sort&compress}

\journal{NIMB}

\begin{document}

\begin{frontmatter}

\title{Measurement of the intrinsic hadronic contamination in the NA64$-e$ high-purity $e^+/e^-$ beam at CERN}
\author[instRussia0]{Yu.~M.~Andreev\orcidlink{0000-0002-7397-9665}}
\author[inst1]{D.~Banerjee\orcidlink{0000-0003-0531-1679}}
\author[inst2]{B.~Banto Oberhauser\orcidlink{0009-0006-4795-1008}}
\author[inst1]{J.~Bernhard\orcidlink{0000-0001-9256-971X}}
\author[inst3,inst3a]{P.~Bisio\orcidlink{/0009-0006-8677-7495}\corref{mycorrespondingauthor}}
\cortext[mycorrespondingauthor]{Corresponding author}
\ead{pietro.bisio@ge.infn.it}
\author[inst4]{M.~Bond\'i\orcidlink{0000-0001-8297-9184}}
\author[inst3]{A.~Celentano\orcidlink{0000-0002-7104-2983}}
\author[inst1]{N.~Charitonidis\orcidlink{0000-0001-9506-1022}}
\author[instRussia0]{A.~G.~Chumakov\orcidlink{/0000-0002-6012-2435}}
\author[inst7]{D.~Cooke}
\author[inst2]{P.~Crivelli\orcidlink{0000-0001-5430-9394}}
\author[inst2]{E.~Depero\orcidlink{0000-0003-2239-1746}}
\author[instRussia0]{A.~V.~Dermenev\orcidlink{0000-0001-5619-376X}}
\author[instRussia0]{S.~V.~Donskov\orcidlink{0000-0002-3988-7687}}
\author[instRussia0]{R.~R.~Dusaev\orcidlink{0000-0002-6147-8038}}
\author[instRussia1]{T.~Enik\orcidlink{0000-0002-2761-9730}}
\author[instRussia1]{V.~N.~Frolov}
\author[inst10]{A.~Gardikiotis\orcidlink{0000-0002-4435-2695}}
\author[instRussia0,inst11]{S.~G.~Gerassimov\orcidlink{0000-0001-7780-8735}}
\author[instRussia0]{S.~N.~Gninenko\orcidlink{0000-0001-6495-7619}}
\author[inst12]{M.~H\"osgen}
\author[inst1]{M.~Jeckel}
\author[instRussia0]{V.~A.~Kachanov\orcidlink{0000-0002-3062-010X}}
\author[instRussia1]{Y.~Kambar\orcidlink{0009-0000-9185-2353}}
\author[instRussia0]{A.~E.~Karneyeu\orcidlink{0000-0001-9983-1004}}
\author[instRussia1]{G.~Kekelidze\orcidlink{0000-0002-5393-9199}}
\author[inst12]{B.~Ketzer\orcidlink{0000-0002-3493-3891}} 
\author[instRussia0]{D.~V.~Kirpichnikov\orcidlink{0000-0002-7177-077X}}
\author[instRussia0]{M.~M.~Kirsanov\orcidlink{0000-0002-8879-6538}}
\author[instRussia0]{V.~N.~Kolosov}
\author[inst11]{I.~V.~Konorov\orcidlink{0000-0002-9013-5456}}
\author[instRussia1]{S.~V.~Gertsenberger\orcidlink{0009-0006-1640-9443}}
\author[instRussia1]{E.~A.~Kasianova}
\author[inst13,inst13aa]{S.~G.~Kovalenko}
\author[instRussia0,instRussia1]{V.~A.~Kramarenko\orcidlink{0000-0002-8625-5586}}
\author[instRussia0]{L.~V.~Kravchuk\orcidlink{0000-0001-8631-4200}}
\author[instRussia0,instRussia1]{ N.~V.~Krasnikov\orcidlink{0000-0002-8717-6492}}
\author[inst13,inst13aa]{S.~V.~Kuleshov\orcidlink{0000-0002-3065-326X}}
\author[instRussia0,inst14,inst13aa]{V.~E.~Lyubovitskij\orcidlink{0000-0001-7467-572X}}
\author[instRussia1]{V.~Lysan\orcidlink{0009-0004-1795-1651}}
\author[inst3]{A.~Marini\orcidlink{0000-0002-6778-2161}}
\author[inst3]{L.~Marsicano\orcidlink{0000-0002-8931-7498}}
\author[instRussia1]{V.~A.~Matveev\orcidlink{0000-0002-2745-5908}}
\author[instRussia0]{Yu.~V.~Mikhailov}
\author[inst13a]{L.~Molina Bueno\orcidlink{0000-0001-9720-9764}}
\author[inst2]{M.~Mongillo\orcidlink{0009-0000-7331-4076}}
\author[instRussia1]{D.~V.~Peshekhonov\orcidlink{0009-0008-9018-5884}}
\author[instRussia0]{V.~A.~Polyakov\orcidlink{0000-0001-5989-0990}}
\author[inst13b]{B.~Radics\orcidlink{0000-0002-8978-1725}}
\author[inst14]{R.~Rojas\orcidlink{0000-0002-6888-9462}}
\author[instRussia1]{K.~Salamatin\orcidlink{0000-0001-6287-8685}}
\author[instRussia0]{V.~D.~Samoylenko}
\author[inst2]{H.~Sieber\orcidlink{0000-0003-1476-4258}}
\author[instRussia0]{D.~Shchukin\orcidlink{0009-0007-5508-3615}}
\author[inst15,inst13aa]{O.~Soto}
\author[instRussia0]{V.~O.~Tikhomirov\orcidlink{0000-0002-9634-0581}}
\author[instRussia0]{I.~Tlisova\orcidlink{0000-0003-1552-2015}}
\author[instRussia0]{A.~N.~Toropin\orcidlink{0000-0002-2106-4041}}
\author[instRussia0]{A.~Yu.~Trifonov}
\author[inst13a]{M.~Tuzi\orcidlink{0009-0000-6276-1401}}
\author[inst13]{P.~Ulloa\orcidlink{0000-0002-0789-7581}}
\author[instRussia0]{B.~I.~Vasilishin}
\author[inst14]{G.~Vasquez Arenas}
\author[instRussia1]{P.~V.~Volkov\orcidlink{0000-0002-7668-3691}}
\author[instRussia0]{V.~Yu.~Volkov\orcidlink{0009-0005-3500-5121}}
\author[instRussia0]{I.~V.~Voronchikhin\orcidlink{0000-0003-3037-636X}}
\author[inst13,inst13aa]{J.~Zamora-Sa\'a\orcidlink{0000-0002-5030-7516}}
\author[instRussia1]{A.~S.~Zhevlakov\orcidlink{0000-0002-7775-5917}}

\address[instRussia0]{ Authors affiliated with an institute covered by a cooperation agreement with CERN}
\address[inst1]{ CERN, European Organization for Nuclear Research, CH-1211 Geneva, Switzerland}
\address[inst2]{ ETH Z\"urich, Institute for Particle Physics and Astrophysics, CH-8093 Z\"urich, Switzerland}
\address[inst3]{ INFN, Sezione di Genova, 16147 Genova, Italia}
\address[inst3a]{ Universit\`a degli Studi di Genova, 16126 Genova, Italia}
\address[inst4]{ INFN, Sezione di Catania, 95125 Catania, Italia}
\address[instRussia1]{ Authors affiliated with an international laboratory covered by a cooperation agreement with CERN}
\address[inst7]{ UCL Departement of Physics and Astronomy, University College London, Gower St. London WC1E 6BT, United Kingdom}

\address[inst10]{ Physics Department, University of Patras, 265 04 Patras, Greece}
\address[inst11]{ Technische Universit\"at M\"unchen, Physik  Department, 85748 Garching, Germany}
\address[inst12]{ Universit\"at Bonn, Helmholtz-Institut f\"ur Strahlen-und Kernphysik, 53115 Bonn, Germany}
\address[inst13]{ Center for Theoretical and Experimental Particle Physics, Facultad de Ciencias Exactas, Universidad Andres Bello, Fernandez Concha 700, Santiago, Chile}
\address[inst13aa]{ Millennium Institute for Subatomic Physics at High-Energy Frontier (SAPHIR), Fernandez Concha 700, Santiago, Chile}
\address[inst13a]{ Instituto de Fisica Corpuscular (CSIC/UV), Carrer del Catedratic Jose Beltran Martinez, 2, 46980 Paterna, Valencia, Spain}
\address[inst13b]{ York University, Toronto, Canada}
\address[inst14]{ Universidad T\'ecnica Federico Santa Mar\'ia and CCTVal, 2390123 Valpara\'iso, Chile}
\address[inst15]{ Departamento de Fisica, Facultad de Ciencias, Universidad de La Serena, Avenida Cisternas 1200, La Serena, Chile}



\begin{abstract}

We present the measurement of the intrinsic hadronic contamination at the CERN SPS H4 beamline configured to transport electrons and positrons at 100 GeV/c. The analysis, performed using data collected by the NA64-$e$ experiment in 2022, is based on calorimetric measurements, exploiting the different interaction mechanisms of electrons and hadrons in the NA64 detector. We determined the contamination by comparing the results obtained using the nominal electron/positron beamline configuration with those from a dedicated setup, in which only hadrons impinged on the detector. We also obtained an estimate of the relative protons, anti-protons and pions yield by exploiting the different absorption probabilities of these particles in matter. We cross-checked our results with a dedicated Monte Carlo simulation for the hadron production at the primary T2 target, finding a good agreement with the experimental measurements.

\end{abstract}


\end{frontmatter}


\section{Introduction: the NA64$-e$ experiment at CERN}

The H4 beamline at the CERN North Area facility is a versatile beamline capable of transporting high-energy particles with momentum in the range of 10-400 GeV/c, with variable composition and purity~\cite{Atherton:164934,Banerjee:2774716}. 
\begin{figure*}[t!]
    \centering
    \includegraphics[width=.8\textwidth]{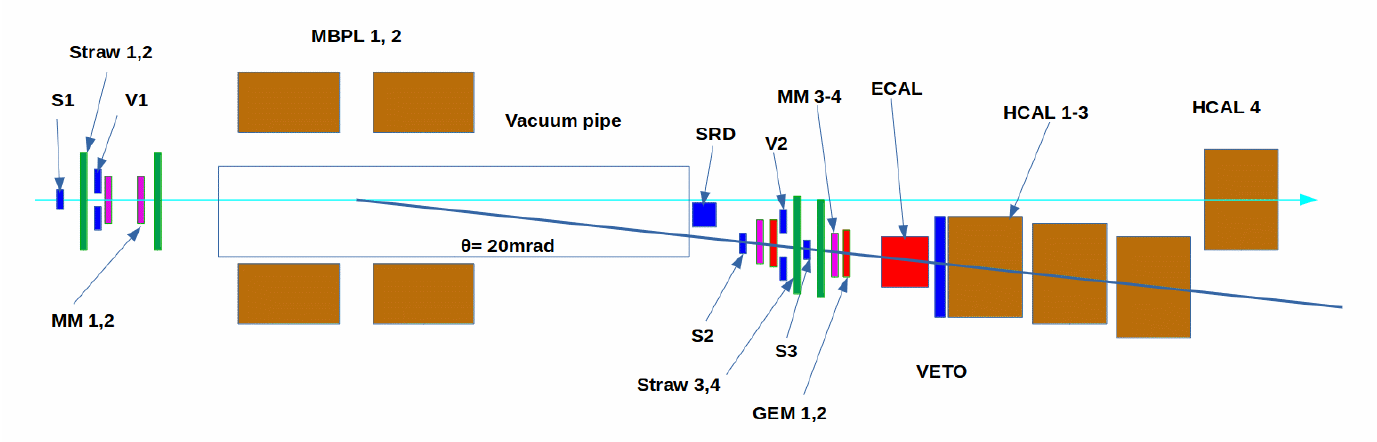}
    \caption{Schematic view of the NA64-$e$ detector in the nominal, invisible mode configuration. See text for further details.}
    \label{fig:detector}
\end{figure*}
The NA64$-e$ experiment exploits the H4 $100$~GeV/c electron beam to search for light dark matter (LDM) particles in the mass range between one and few hundred MeVs~\cite{Fabbrichesi:2020wbt,Filippi:2020kii,Graham:2021ggy}. NA64 uses the ``missing-energy'' technique: the beam collides on an active target measuring the  energy deposited  by  each impinging particle; LDM particles produced by the interaction of the primary electron with the active target escape from the latter undetected. The signal signature is the observation of events with a large \textit{missing energy}, defined as the difference between the nominal beam energy and the one deposited in the target. Beam purity is a critical parameter for the experiment, since contaminating hadrons are a potential source of background events~\cite{NA64:2019imj}. 

The NA64$-e$ detector is schematized in Fig.~\ref{fig:detector}. It consists of (I) a magnetic spectrometer to measure the momentum of each impinging particle, made by two successive dipole magnets and a set  of tracking detectors -- Micromegas, GEMs, and Strawtubes~\cite{Volkov:2019qhb} -- installed upstream and downstream the magnet, (II) a syncrotron radiation beam-tagging system (SRD) based on a Pb/Sc sandwhich calorimeter detecting the SR photons emitted by the electrons due to their bending in the dipole magnetic field,~\cite{Depero:2017mrr}, (III) a 40-radiation length electromagnetic calorimeter (ECAL), serving as active thick target, with energy resolution $\sigma_{E}/E\simeq 10\%/\sqrt{E\mathrm{(GeV)}}\oplus4\%$, (IV) a high-efficiency plastic scintillator counter (VETO) used to identify charged particles produced  by the interaction of the primary beam with the ECAL, and (V) a downstream massive and hermetic hadronic calorimeter used to detect secondary long-lived neutral hadrons such as neutrons and $K_L$ (HCAL). The ECAL is assembled as a $5\times6$ matrix of $3.82\times3.82$ cm$^2$ cells with independent PMT readout, segmented into a $4X_0$ pre-shower section (PS) and a main section. Overall, its lenght corresponds to approximately 1.3 nuclear interaction lengths.
The HCAL length corresponds to $\simeq21$ hadronic interaction lengths, resulting in a punch-through probability of about $10^{-9}$. A fourth HCAL module (HCAL-0) is installed at zero degrees to measure neutral hadrons produced by upstream interactions of the primary beam with the beamline elements\footnote{In the following, we'll denote as ``HCAL'' the combination of the three modules installed downstream the ECAL.}. The production trigger for the experiment requires the coincidence between the signals of a set of upstream beam-defining plastic-scintillator counters (SC), as well as an in-time cluster in the ECAL with total energy $E_{ECAL} \lesssim  80$~GeV and pre-shower energy $E_{PS}\gtrsim 300$~MeV. For calibration purposes, an ``open-trigger'' is also implemented, requiring the coincidence between the SC signals solely. Data used for the study presented in this work were acquired with the nominal detector configuration described before, in ``open-trigger'' mode. 

The NA64$-e$ experiment imposes strict requirements on the properties of the impinging beam. The beam current should be low enough to allow to resolve each individual electron/positron impinging on the detector, allowing at the same time to accumulate a large statistics: ideally, an impinging particle rate of about $1\mbox{--}10$ MHz is required. Furthermore, the intrinsic beam energy distribution should be as narrow as possible, to allow for a proper measurement of the missing energy. Considering the nominal ECAL energy resolution for a 100 GeV impinging beam of about $3\mbox{--}4\%$, the required beam energy spread should be of about $1\mbox{--}2\%$ or lower. Finally, the NA64$-e$ missing-energy trigger condition reflects on the maximum allowed intrinsic hadronic contamination of the beam. 

During operations, two main types of events are recorded. The first is associated with the production of energetic particles escaping from the active target by the interaction of the primary $e^-/e^+$ or one of its secondaries with the ECAL. Events of this first kind are, for example, the electro-production of energetic hadrons, as well as the radiative production of a forward muon pair (so-called ``di-muon production''); the LDM signal also enters in this category. The second source of measured events is due to the interaction with the target of beam hadron contaminants, with only partial deposition of the primary beam energy in the ECAL. 

A detailed knowledge of these two event sources is required to tune the trigger thresholds and evaluate the corresponding performances. While the first source of events can be efficiently studied by means of Monte Carlo simulations, a proper control and estimate of the second requires a detailed knowledge of the intrinsic hadronic contamination of the primary beam impinging on NA64$-e$. In this work, we present the results obtained from a dedicated measurement of the intrinsic hadronic contamination affecting the electron/positron beam from the H4 beamline performed with the NA64 detector.

\section{The H4 beamline at CERN North Area}

The H4 beam is obtained by having a primary 400 GeV/c proton beam from the Super Proton Synchrotron (SPS) accelerator impinging on a thin beryllium target, and then selecting secondary or tertiary particles by means of a set of magnets and beam absorbers/attenuators~\cite{Brianti:604383,Booth:2019brj}. The particles produced at the target are momentum-selected and transported through a $\simeq$ 540-m long beamline, composed of many bending dipoles, focusing quadrupoles and corrector elements towards the experimental area. Collimating structures and beam instrumentation are also present and used in order to ensure the beam properties on a spill by spill basis.

\subsection{Electron/positron beam production: the T2 target}

The production target serving the H4 (and H2) beamlines (designated ``T2'' target) is a 500-mm long Be plate, with transverse size 160 mm (horizontal) $\times$ 2 mm (vertical), where the 400 GeV/c $\pm 0.3\% ~ \frac{\delta p}{p}$ proton beam is slowly extracted on~\cite{Booth:2019brj,Fraser:2019mpj}. The intensity per unit time of the protons incident on the T2 target varies depending on the other SPS operations;
typical values are of the order of about $2\mbox{--}3 \times 10^{12}$ protons per 4.8 s spill, with one or two spills per supercycle. The supercycle length varies between 14.4 and 60 seconds. The target position with respect to the primary SPS beam direction, as well as the configuration of the selection dipoles and the beam absorbers/attenuators depends on the secondary beam to be produced and delivered to the experimental area.

\begin{figure*}[t]
    \centering
    \includegraphics[width=.95\textwidth]{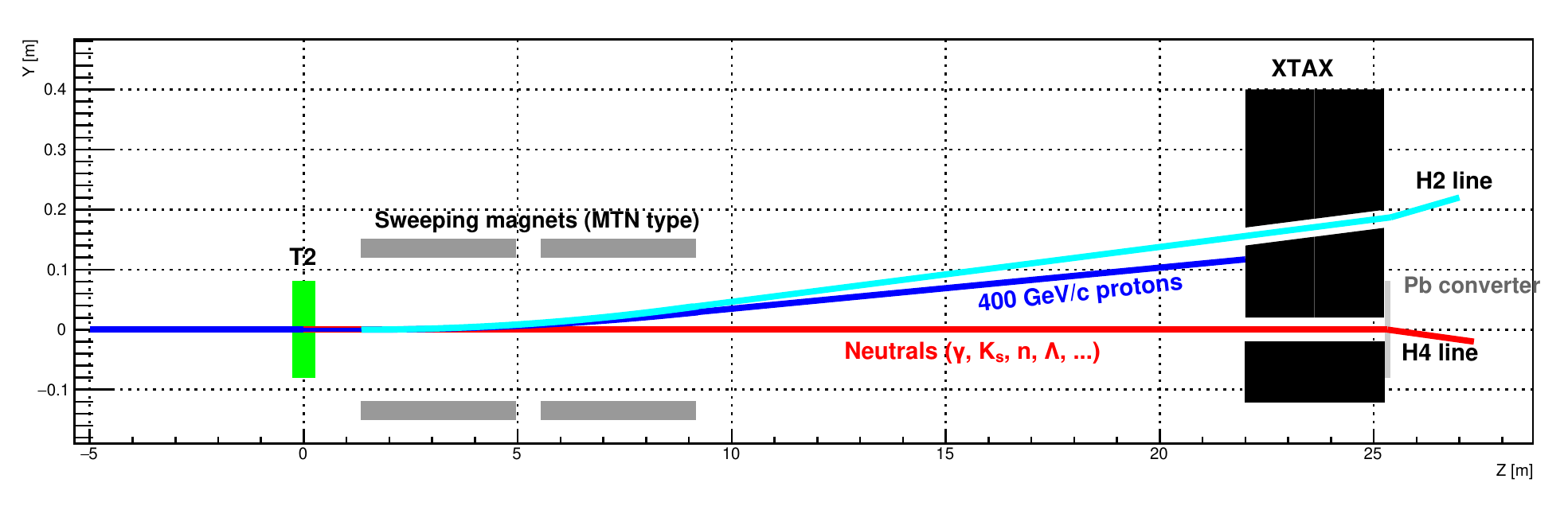}
    \caption{A schematic representation of the main beamline elements after the T2 target. The neutral particles are going straight through the XTAX hole and converted in the converter which is just after. The grey horizontal lines correspond to the magnet apertures, the blue line is the 400 GeV/c beam while the red solid line corresponds to the trajectory of the neutral particles before they impinge on the converter. The black structure on the right part of the figure correspond to the XTAX apertures, as discussed in the text. A comprehensive description of the H4 beamline optics is presented in Ref.~\cite{Banerjee:2774716}.}
    \label{fig:beam_production}
\end{figure*}

When operated in the special electron/positron mode for the H4 line, leptons are produced via a dual conversion process, by having the decay photons from $\pi^0/\eta$ mesons produced in the target propagating downstream at zero production angle and pair-converting on a thin lead target installed downstream. In general, for a given electron / positron energy, the yield is governed by the integral of $\pi^{0}$ decays that lie above the momentum considered, while it rapidly decreases with increasing production angle. 
A simplified drawing of the T2 target station elements in this special configuration is shown in Fig.~\ref{fig:beam_production}. Downstream the target, two large aperture bending dipole magnets (MTN) are installed, each with a length of 3.6~m~\cite{Keizer:319359}. The end of the first (second) is located at 4.95~m (9.15~m) from the center of the target, with a 0.6~m drift volume in between them. The scope of these magnets is to sweep away all secondary charged particles produced in the T2 target and also deflect the 400 GeV/c beam on a downstream absorber (XTAX). The magnetic field of each MTN is directed vertically, while the strength is regulated to have the SPS proton beam being deflected in the horizontal plane by an angle of 6.85~mrad -- for a primary momentum of $400$~GeV/c, a total magnetic field integral $\int \vec{B}\cdot d\vec{l}=4.57$~T$\cdot$m is required for each. The useful aperture of the two magnets is 240 mm $\times$ 60 mm.
The XTAX is made of two large collimating structures, consisting of 1.615-m thick massive blocks constructed mainly from stainless steel, with the end of the first (second) located at 23.615~m (25.240~m) from the center of the target. The XTAX has several holes with different diameters, vertically aligned, to allow the passage of the secondary particles of interest. 
In the case of electron/positron beam configuration, a 64$\times$64 mm$^2$ hole is aligned with the primary beam direction before the target. The secondary target for pair production is a 4~mm thick lead converter, located at 25.323~m from the T2 center. After the converter, at the start of H4 beamline, another horizontally-deflecting septum magnet with a length of 3.2 m and aperture 114$\times$ 60 mm$^2$, whose end is located at 28.850~m from the target, is used to perform a first momentum and charge sign selection of the particles that are transported to the experimental area. 

\subsubsection{Hadronic contaminants} \label{subsec:Hadronic contaminants}
In the electrons/positrons configuration, the main source of hadron contaminants in the beam is the forward production of long-lived neutral particles in the target, such as $\Lambda$ hyperons and $K_S$, propagating downstream and decaying to charged particles after the sweeping magnet. If secondary particles produced at the XTAX, the vacuum chambers, the surrounding shielding or even the subsequent septum magnet aperture are within the proper momentum, spatial and angular acceptance, they could be transported by the H4 beamline towards the experimental area. However, most of these particles only make it up to the 
section of the line where a momentum selection of $p_{0} \pm 1.2\% $ (maximum) takes place, filtering out all particles outside this very narrow momentum band. This selection, combined with synchrotron radiation effects (present in higher momenta) essentially make the beams reaching the experimental areas very pure (typically above 90\%).

\begin{figure}
    \centering
    \includegraphics[width=.49\textwidth]{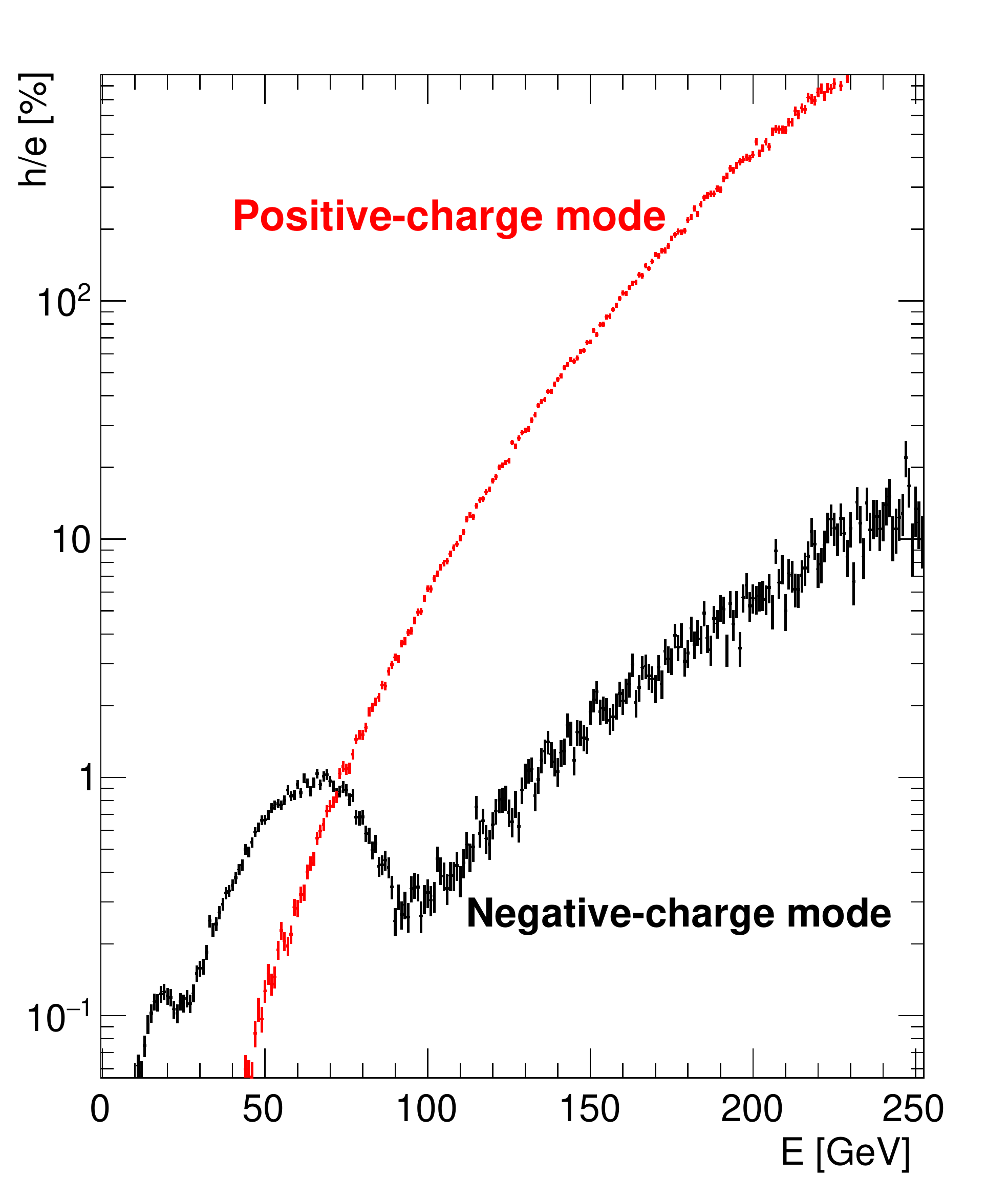}
    \caption{The FLUKA calculated ratio between hadrons and electrons / positrons at the H4 lead converter. The angular and momentum acceptance of H4 beamline have been applied both for the negative-charge (black) and positive-charge (red) mode. The structure at $E\simeq50$ GeV for the negative charge mode is a result of the convolution between the energy spectrum of the produced $\Lambda$ baryons and the maximum energy allowed in the $\Lambda \rightarrow p \pi^-$ decay. }
    \label{fig:T4_sim1}
\end{figure}

When the beamline is operated in negative-charge mode ($e^{-}$), the contamination in the low momentum range, $p_0\lesssim 100$~GeV/c, is mostly due to the pions from $\Lambda \rightarrow p\pi^-$ decay. At larger momentum this contribution drops because of the kinematical limit of the decay process\footnote{Starting from a $p_0=400$~GeV/c proton beam, the maximum energy of the $\pi^-$ from the decay of a $\Lambda$ baryon produced in the Be target is $E^\pi_{max}\simeq \frac{p_0}{M_\Lambda} \cdot (E^*_\pi+p^*_\pi)$, where $E^*_\pi$ ($p^*_\pi$) is the pion energy (momentum) in the $\Lambda$ rest frame. Numerically, $E^\pi_{max}\simeq97$~GeV.  The proton maximum energy is $E^p_{max}\simeq375$~GeV.
For comparison, the maximum pion energy from the $K_s\rightarrow \pi^+ \pi^-$ decay is
$E^\pi_{max}\simeq\frac{p_0}{M_K}(E^*_\pi+p^*_\pi)$. Numerically, $E^\pi_{max}~\simeq366$~GeV.}, and the main source of hadron contaminants is the $K_S$ decay to a $\pi^+\pi^-$ pair. Residual contributions are due to anti-protons from $\overline{\Lambda}\rightarrow \overline{p}\pi^+$ decay, as well as from prompt charged particles produced in the T2 target at non-zero angle and then re-deflected by the MTN magnets toward the XTAX hole and the converter. In positive-charge mode ($e^{+}$), instead, there is no kinematic suppression at large momentum for the protons from $\Lambda$ decay. Therefore, a larger intrinsic hadronic contamination of the beam is expected with respect to the electrons one, due to the much smaller $\bar{\Lambda}$ yield. This effect is illustrated in Fig.~\ref{fig:T4_sim1}, showing the H4 beam hadrons-to-electrons ($h/e$) ratio as a function of the energy in the negative-charge (black) and positive-charge (red) mode, as obtained from a FLUKA-based simulation\footnote{We used the \texttt{PRECISIO} default settings.}~\cite{Ferrari:2005zk,Bohlen:2014buj}. In the simulation, we included the T2 target, the dipole sweeping magnets, the XTAX, and the lead conversion target. We computed the $h/e$ ratio by sampling all particles emerging from the latter. To account for the acceptance of the H4 beamline, we imposed the following kinematic cuts: $\left|p_x/p\right| < 1\%$, $\left |p_y/p \right| < 1\%$, $\left |x_T \right| < 5$~mm, $\left | y_T \right|<5$~mm, where $p_x$ ($p_y$) is the particle momentum in the horizontal (vertical) direction, $p$ is the total momentum, and $x_T$ ($y_T$) is the horizontal (vertical) coordinate of the particle position at the target center, obtained by projecting straight back from the converter to the T2 target center. At 100 GeV, the hadron contamination in negative-charge mode is of about 0.2-0.3$\%$, while for the positive-charge mode is roughly one order of magnitude higher.

\begin{figure*}[t]
    \centering
    \includegraphics[width=.95\textwidth]{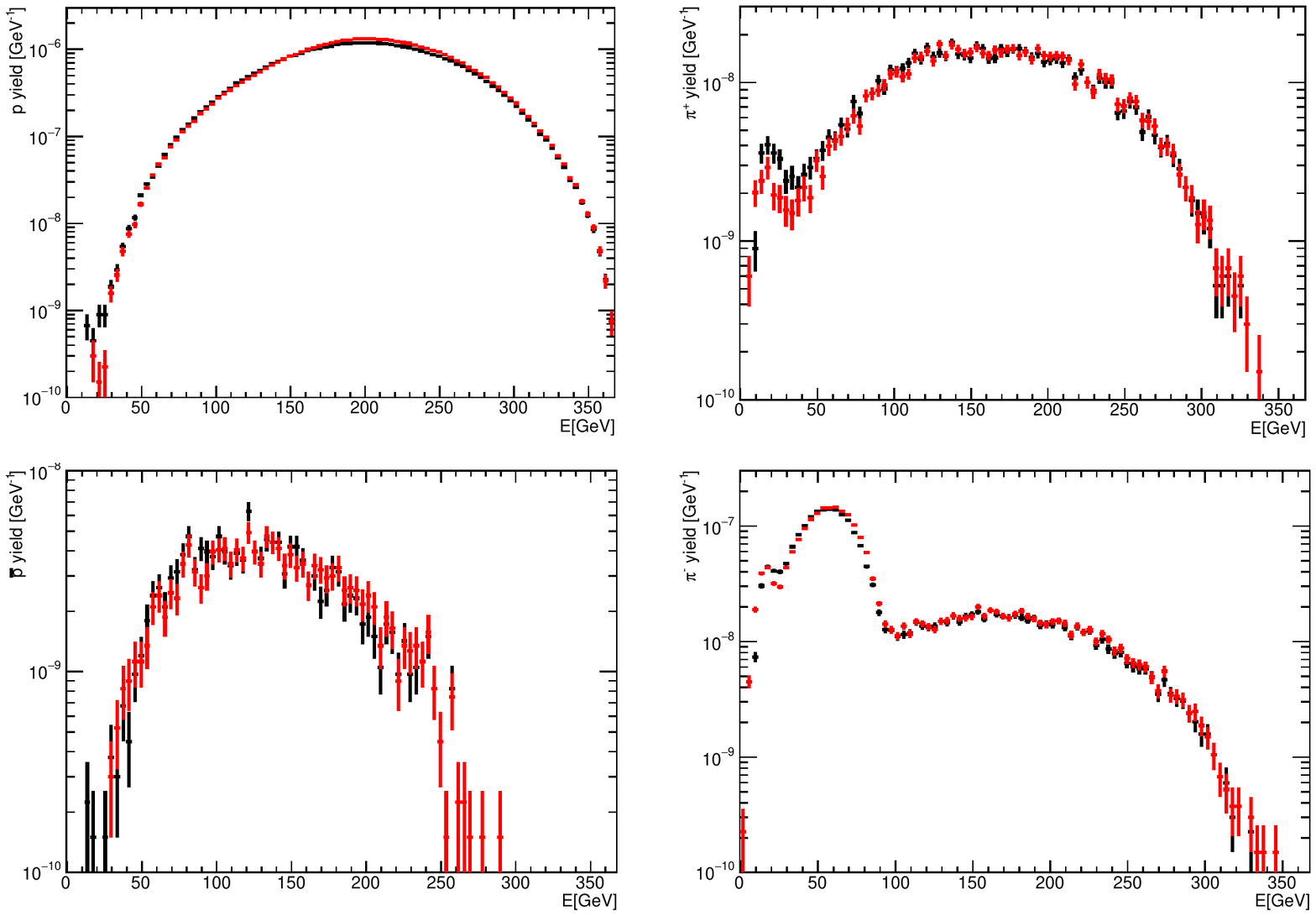}
    \caption{The differential yield of protons (top-left), $\pi^+$ (top-right), anti-proton (bottom-left), and $\pi^-$ (bottom-right) after the Pb conversion target obtained from FLUKA, including (black) or not (red) the latter in the simulation. All results have been normalized to the total number of impinging protons on the T2 target.}
    \label{fig:T4_sim2}
\end{figure*}

Finally, we observe that a residual background source is associated with the photo-production of heavy charged particles in the converter. For example, muons can be radiatively produced from the process $\gamma Pb \rightarrow \mu^+\mu^- Pb$. However, the cross-section for this reaction is suppressed by a factor $\left(\frac{m_e}{m_\mu}\right)^2\simeq 2.2 \cdot 10^{-5}$ with respect to $e^+e^-$ pair production, making this negligible. Similarly, to get a first estimate of the charged hadrons photo-production, we assume a total $\gamma-p$ hadronic cross section at $E_\gamma \simeq 100$~GeV of $\sigma_{\gamma p}\simeq 200$~$\mu$barn, and the simple incoherent scaling relation $\sigma_{\gamma Pb}\simeq A \sigma_{\gamma p}$, where $A$ is the atomic number. This results to a total number of hadronic interactions of about $4\cdot 10^{-4}$ per impinging photon on the converter, to be compared to the fraction of photons undergoing an $e^+e^-$ pair conversion of about $s_{Pb}/X_0~\simeq 1$. In conclusion, the photo-production of heavy charged particles from the converter is negligible with respect to the decay mechanisms previously discussed. This is also highlighted by the energy spectra reported in Fig.~\ref{fig:T4_sim2}, comparing the results obtained including (black) or not (red) the lead conversion target in the FLUKA simulation.

\section{Methodology}

\subsection{Experimental setup and data sample}

We measured the H4 $e^-/e^+$ beam intrinsic hadronic contamination by using data collected by NA64 in 2022 during so-called ``calibration runs''. In these data takings, the experiment operated in ``open-trigger'' mode. Data were acquired  with and without the lead converter after the XTAX. In runs performed using the converter (``electron calibration runs''), the beam impinging on the detector is composed of electrons and of a small fraction of contaminating hadrons. The measurement of this hadronic contaminants fraction is the ultimate goal of this work. On the other hand, in runs performed without the converter (``hadron calibration runs''), the beam is almost entirely composed of hadrons. In both configurations, a small fraction of muons, produced by pion decay, is also present. 

Overall, we collected eight pairs of negative-charge calibration runs and one in positive-charge mode (see table~\ref{table:runs}). Each pair consists of a run performed without the lead converter and one with, acquired in series, not changing any other beamline configuration. This procedure guarantees that the data measured in the first run are representative of the hadronic contamination in the second one. The typical beam intensity in the electron calibration runs (converter installed) was about $6 \sim 10^{6}$ particles/spill.

\begin{table}[b]
\centering

\begin{tabular}{c c c c}\hline\hline
H4 config. & Period & Charge & h/e (\%) $\pm$ Stat $\pm$ Syst \\
\hline
\setlength\tabcolsep{2pt}
(a)&$\left\{ \begin{tabular} {c}
I\\
II
\end{tabular}\right.$ 
&
\begin{tabular}{c}
Negative\\
Negative
\end{tabular}
&
\begin{tabular}{c}
0.313 $\pm$ 0.015 $\pm$ 0.002 \\
0.323 $\pm$ 0.015 $\pm$ 0.001
\end{tabular}
\\
(b) & $\,\,$III & Negative & 0.356 $\pm$ 0.017 $\pm$ 0.002 \\
\setlength\tabcolsep{1.5pt}
(c) &$\left\{ \begin{tabular}{c}
IV\\
V\\
VI
\end{tabular}\right.$ 
&
\begin{tabular}{c}
Negative\\
Negative\\
Negative
\end{tabular}
&
\begin{tabular} {c} 
0.380 $\pm$ 0.017 $\pm$ 0.002 \\
0.386 $\pm$ 0.015 $\pm$ 0.002\\
0.389 $\pm$ 0.012 $\pm$ 0.001
\end{tabular}
\\
\setlength\tabcolsep{1.5pt}
(d)&$\left\{\begin{tabular}{c}
VII\\
VIII
\end{tabular}\right.$ 
&
\begin{tabular}{c}
Negative\\
Negative
\end{tabular} 
&
\begin{tabular}{c}
0.389 $\pm$ 0.012 $\pm$ 0.001\\
0.367 $\pm$ 0.016 $\pm$ 0.002
\end{tabular}
\\
(e)&$\,\,\,\,$IX & Positive & 4.29\hspace{2.7mm}$\pm$ 0.09\hspace{2.7mm}$\pm$ 0.009 \\ \hline
\end{tabular}

\caption{The table shows the run pairs analyzed in this study with the corresponding charge configuration. The intrinsic hadronic contamination, measured as described in the text, is reported here, together with the evaluation of statistical and systematic uncertainty. We grouped together runs referring to the same H4 beamline configuration (collimators opening), as described in the text.} 
\label{table:runs}
\end{table}

\subsection{Data analysis}\label{subsec:Data analysis}

Our analysis is based on the methodology summarised below. First, exploiting hadron calibration runs, we evaluated the fraction $f$ of impinging hadrons that interact with the ECAL only through electromagnetic ionisation and deposit all their energy ($\sim$ 100 GeV) in the HCAL. Subsequently, through the same procedure, we determined in electron calibration runs the total number of events with the same topology. Assuming that $f$ is the same in both run modes, we could finally extract the total number of hadrons and thus determine the relative hadron contamination $h/e$, where $e$ ($h$) is the fraction of electrons (hadrons) in the beam.

\begin{figure*}
    \centering
    \includegraphics[width=.49\textwidth]{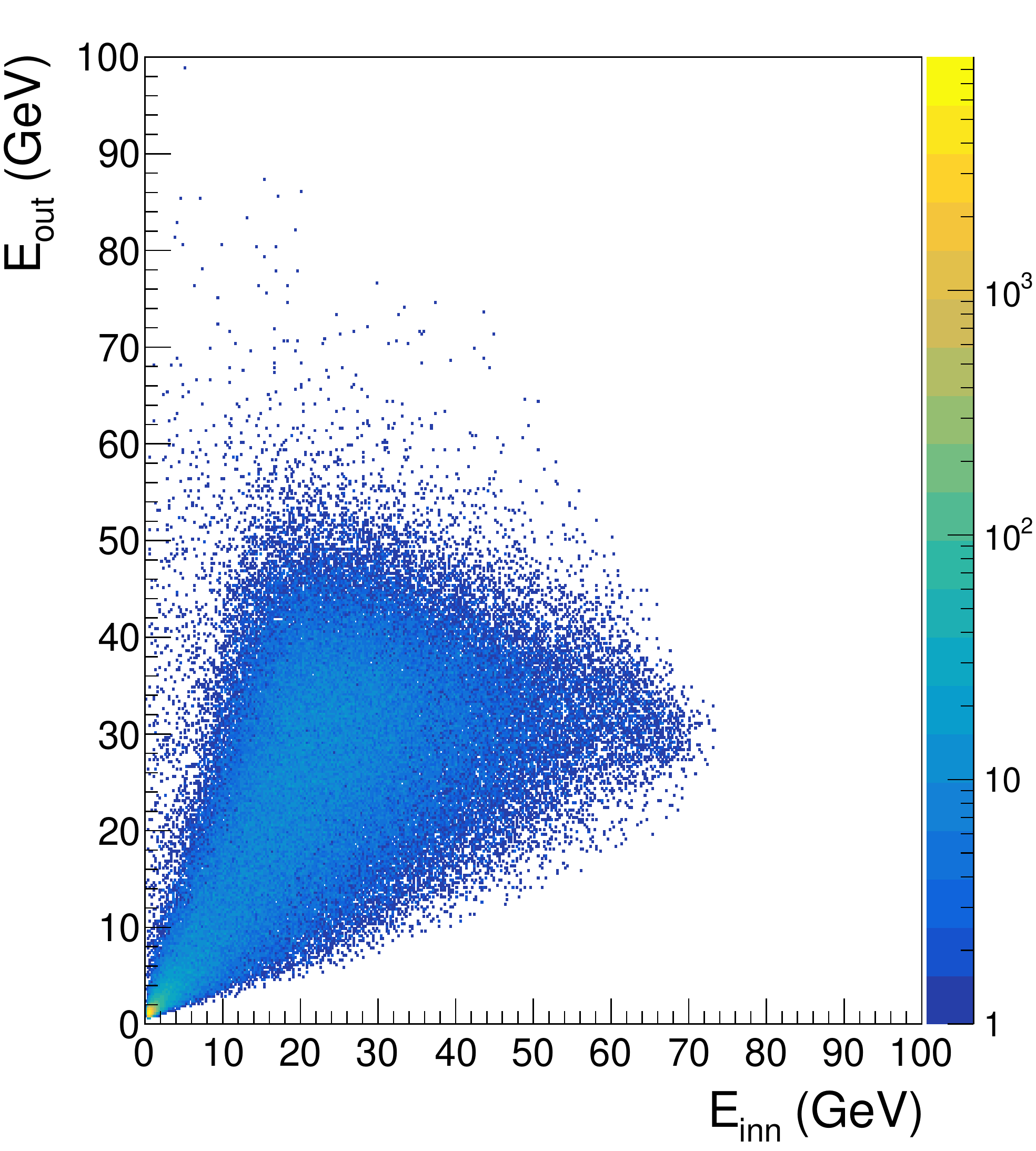}
    \includegraphics[width=.49\textwidth]{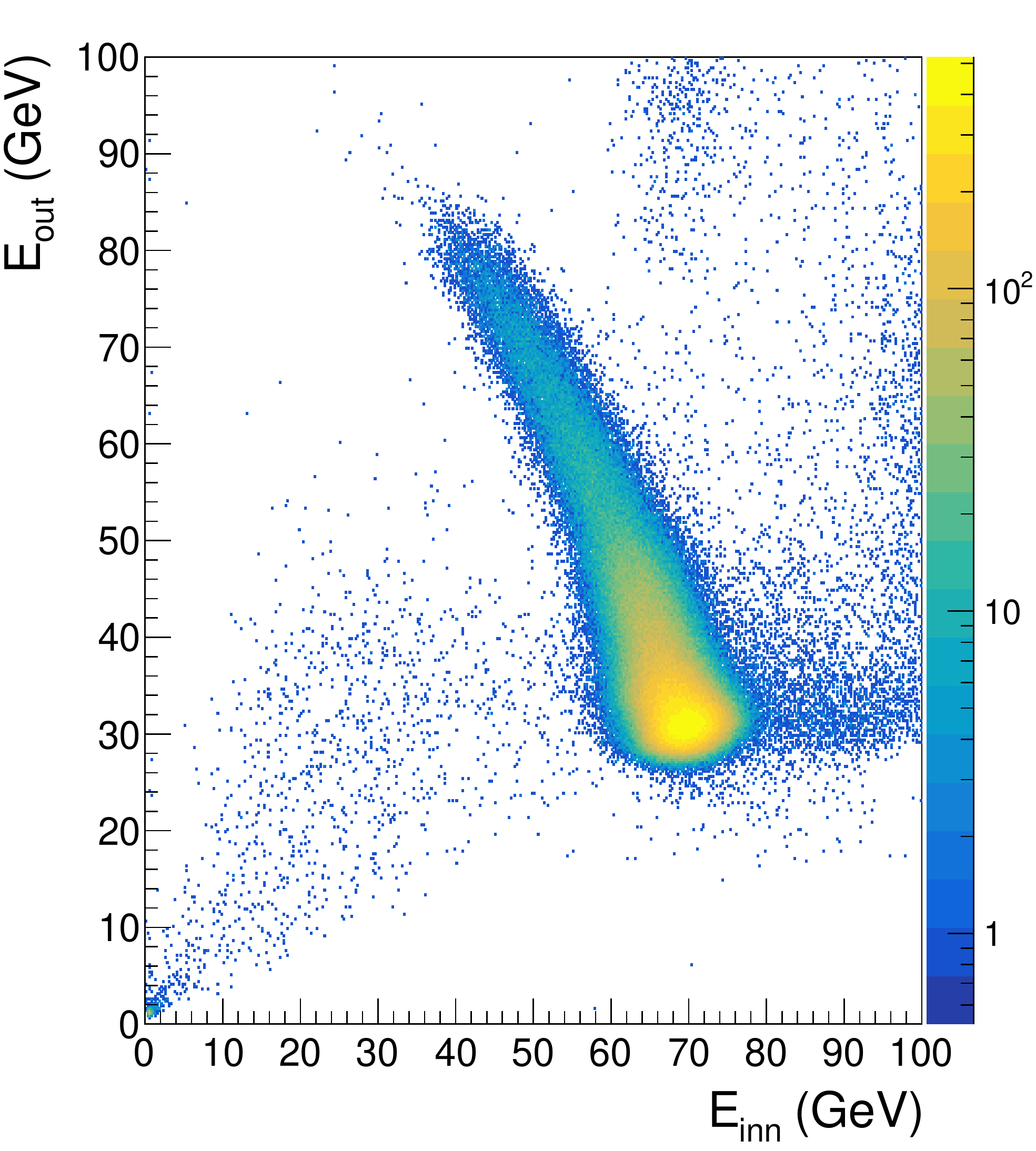}
    \caption{The $E_{out}$ VS $E_{inn}$ distribution for a hadron (left) and electron (right) calibration run. In the right plot, the presence of contaminating hadrons manifests as the events accumulating in the low-$E_{out}$ low-$E_{inn}$ energy region, generating a "triangular" distribution similar to that reported in the left plot. }
    \label{fig:innVSout}
\end{figure*}

\begin{figure}
    \centering
    \includegraphics[width=.49\textwidth]{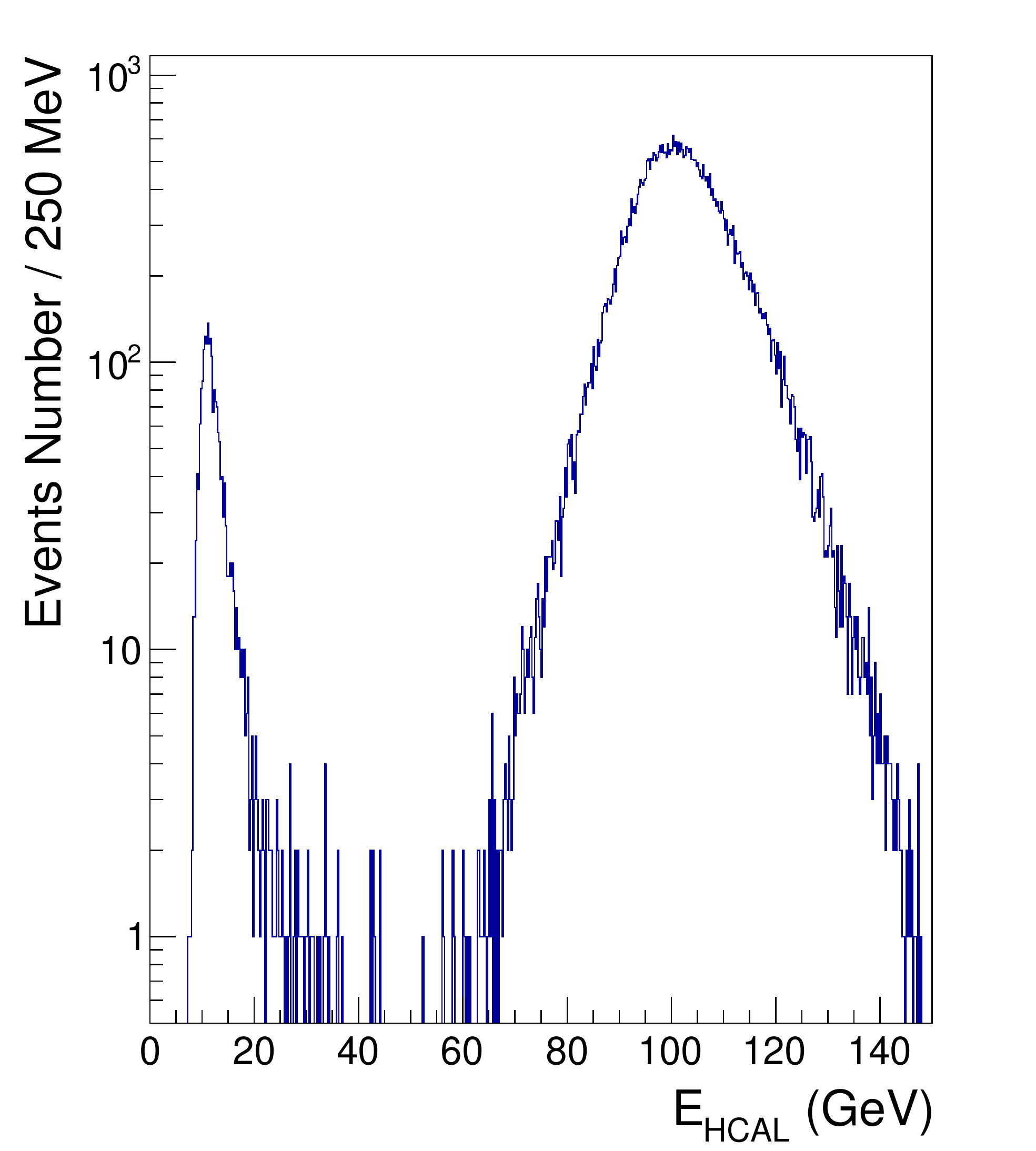}
    \caption{The $E_{HCAL}$ distribution for a hadron calibration run after the $S_{E}$ MIP-like events selection was applied. The low-energy peak at $E_{HCAL}\simeq 10$ GeV is due to passing-through muons, while the high-energy peak is due to 100 GeV fully-absorbed hadrons.}
    \label{fig:6895_HCAL}
\end{figure}

To evaluate $f$, we first selected particles that act as MIPs within the ECAL ($S_E$ selection), by applying a threshold on the energy deposited in the ECAL central cell: $E_{inn} < 5 $ GeV. At the same time, we required that the energy $E_{out}$ deposited in all other cells was less than 7 GeV. The observed $E_{out}$ vs $E_{inn}$ distribution is reported in Fig.~\ref{fig:innVSout} for an electron run and a hadron run. These histograms evidence the different topologies of events caused by hadrons and electrons. After the $S_{E}$ MIP-like events selection, we applied a cut on the total energy deposition in the HCAL to distinguish hadrons from muons, $E_{HCAL}>50$~GeV (this selection is referred to as $S_{H}$). Figure~\ref{fig:6895_HCAL} reports the $E_{HCAL}$ distribution for a hadron calibration run, showing two distinct peaks. The low-energy peak is due to events in which a muon impinges on the NA64 setup and passes through the calorimeters depositing a small amount of energy due to ionization. The high-energy peak is instead due to hadrons entirely absorbed in the HCAL.

The fraction $f$ was measured from hadron calibration data as the ratio between the number of events satisfying the $S_{E} + S_{H}$ selection $N^{h}_{S_{E} + S_{H}}$, and the total number of events $N^{h}$:
\begin{equation}\label{eq:f}
    f = \frac{N^{h}_{S_{E} + S_{H}}}{N^h} \; \;.
\end{equation}
Subsequently, we determined the number of events satisfying the $S_{E} + S_{H}$ selection for electron calibration data, $N^{e}_{S_{E} + S_{H}}$, and converted it to the total number of hadrons normalizing by $f$. The $h/(h+e)$ ratio thus reads:
\begin{equation}\label{eq:h/e}
    \frac{h}{h+e}=\frac{N^{e}_{S_{E} + S_{H}}}{f}\frac{1}{N^{e}} \; \;,
\end{equation}
where $N^{e}$ is the total number of events collected with the lead converter. This procedure was applied independently for each run pair.

In the above formulas, all particle yields must be corrected to account for the presence of muons in the beam. The corresponding correction factors were determined via Monte Carlo simulations, as described in the following subsection.

\subsection{Monte Carlo simulations}

We simulated about $5 \times 10^6 $ muons impinging on the detector using the official NA64 simulation software, based on the Geant4 framework~\cite{Agostinelli:2002hh,Allison:2006ve}. According to the described selection, the events caused by muons can be divided into three distinct categories. Most muons pass through both ECAL and HCAL interacting solely through ionization and depositing a small amount of energy in both detectors. This class of events results in a clear signature and satisfies the $S_{E}+\overline{S}_{H}$ selection. We estimated the corresponding relative fraction to be $f^\mu_1\simeq 98.4\%$. The second class of events is due to muons crossing the ECAL and depositing more than 50 GeV in the HCAL, satisfying the $S_{E}+S_{H}$ selection and mimicking the hadron behaviour. We investigated the nature of these events and found them to be typically characterized by the emission of a high-energy bremsstrahlung photon interacting with the HCAL. We estimated the corresponding relative fraction to be $f^\mu_2\simeq 0.8\%$. The last category of events includes those in which the muon gives rise to a large energy deposition in the ECAL and thus does not satisfy the $S_{E}$ selection. A deeper scrutiny of these events showed that they are mostly associated with an intense ionization ($\delta-$ray emission) in the calorimeter. The corresponding fraction of events was found to be $\simeq 0.8\%$. 

Similarly, we simulated the interaction of pions with the NA64 detector, starting from about $5 \times 10^6 $ Monte Carlo events. In particular, we focused on events with a MIP-like signature in the ECAL (i.e. passing the $S_{E}$ selection) and a small energy deposition in the HCAL (not verifying condition $S_{H}$), and found them to be $\simeq 0.3\%$. However, simulations show that in all these events the primary pion decays within the beamline before reaching the target, generating a high-energy muon impinging on the ECAL; this is compatible with the observation that, given the large thickness of the HCAL ($\sim 30 \,\lambda_I$), the probability for a pion to pass through it without any hard interaction is completely negligible.

In conclusion, simulations show that the low-energy peak in the HCAL spectrum is solely populated by events caused by impinging muons. Therefore, in experimental data, we considered that all events satisfying the $S_E+\overline{S}_H$ selection are originated by these particles. For each run we could thus estimate the total number of impinging muons from the following equation: 
\begin{equation}
    N^{\mu}=\frac{N^{h/e}_{S_E+\overline{S}_H}}{f^\mu_1} \; \; ,
\end{equation}
and then subtract this yield from the two denominators $N^h$ (Eq.~\ref{eq:f}) and $N^e$ (Eq.~\ref{eq:h/e}). Similarly, the two terms $N^{h}_{S_{E} + S_{H}}$ and $N^{e}_{S_{E} + S_{H}}$ were corrected by subtracting the quantity $ N^{\mu} \cdot f^\mu_2$.

\section{Results}\label{sec:Results}

Using the technique described above, we could estimate the hadron contamination for the different runs studied. The obtained measurements are reported in table~\ref{table:runs}. We quoted the error deriving from the statistical uncertainty on the measured particle yields, with the major contribution being that from the two terms $N^h_{S_E+S_H}$ and $N^e_{S_E+S_H}$. 

\begin{figure}
    \centering
    \includegraphics[width=.49\textwidth]{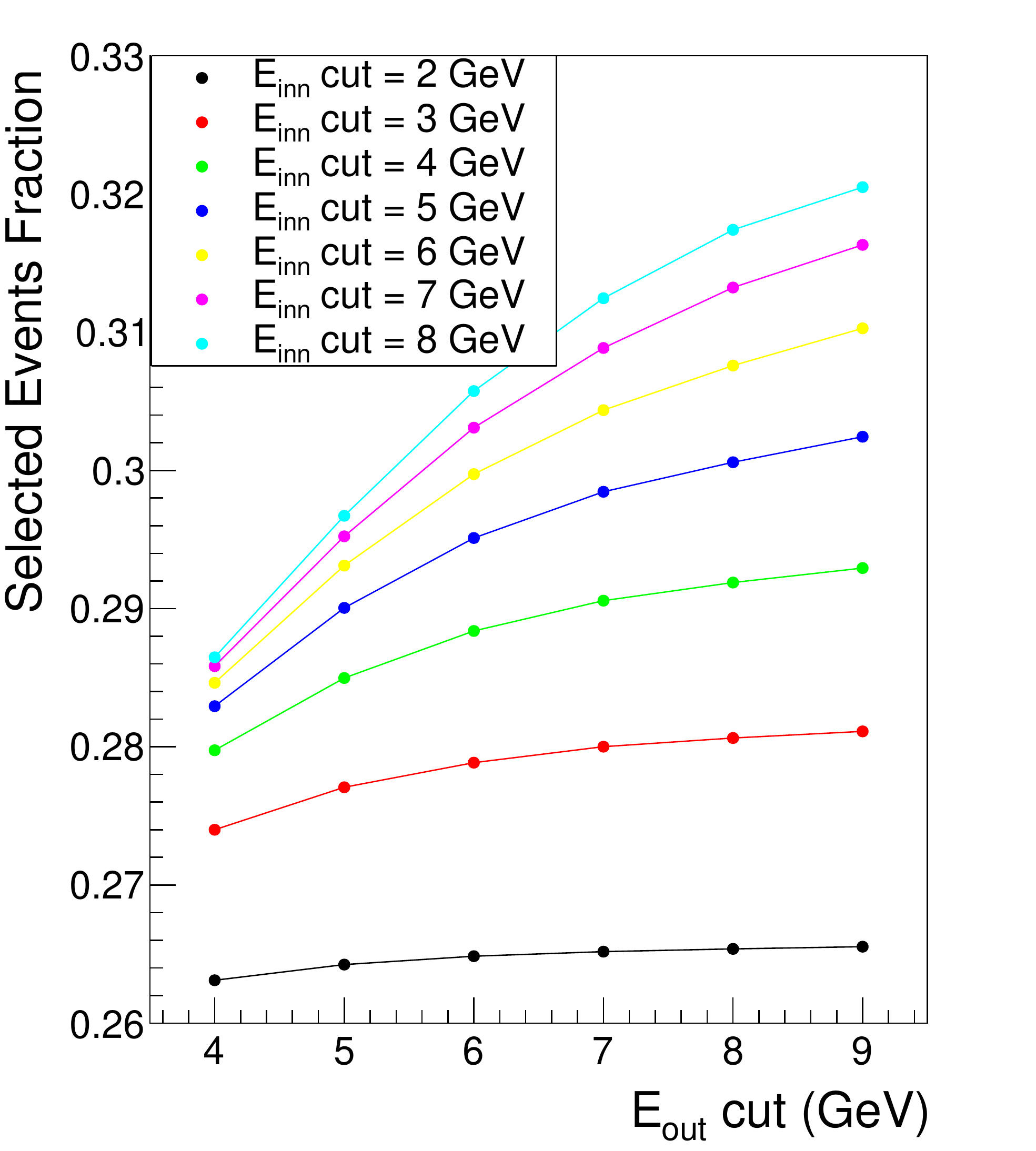}
    \caption{The fraction of hadron calibration run events passing the $S_E$ selection as a function of the $E_{inn}$ and $E_{out}$ cut thresholds.}
    \label{fig:6895_SE}
\end{figure}

To determine the systematic uncertainty of our results, we performed a dedicated study evaluating the effect of varying the thresholds defining the selection of events. In particular, for $S_E$, we modified the cut on $E_{inn}$ from 2 GeV to 8 GeV in steps of 1 GeV. In the same way, we varied the threshold on $E_{out}$ from 4 GeV to 10 GeV. For each combination, we repeated the evaluation of $f$ and $h/e$. Although the value of $f$ significantly depends on the selection thresholds, as shown in Fig.~\ref{fig:6895_SE}, the hadronic contamination $h/e$ is affected very little by these variations. This result is due to the applied methodology, which founds on the relative ratio of events selections in hadron and electron runs, and the variations on $f$ factorize out. For $S_H$, since the muon and the hadron populations are clearly distinct in the HCAL energy distribution, we noticed that no variations for these observables were induced by changing the corresponding cut values. Using this approach for each run pair, we evaluated the systematic uncertainty as the standard deviation of $h/e$ measurements obtained for different cut combinations. In conclusion, this study determined that the systematic uncertainty affecting $h/e$ is negligible compared with the statistical one, being smaller by a factor $\sim10$. We quote and report our results in table~\ref{table:runs}. 

The obtained results agree with the simulations' predictions reported in Fig.~\ref{fig:T4_sim1}. In particular, the $h/e$ ratio, for the negative charge configuration runs is $\sim0.3\mbox{--}0.4~\%$, in good agreement with the result of the simulations. Similarly, for the positron run, we estimated that the hadron contamination was about $\sim4~\%$, to be compared with the $\sim 6~\%$ prediction from Monte Carlo. However, we should note that the Monte Carlo simulation does not include the $\simeq 540$ m beamline between the T2 target and the NA64 detector. This transport alters the $h/e$ ratio because of the synchrotron radiation effects and of the pion decay that reduces the fraction of hadron transported through the line, increasing the population of muons.

\begin{figure}
    \centering
    \includegraphics[width=.49\textwidth]{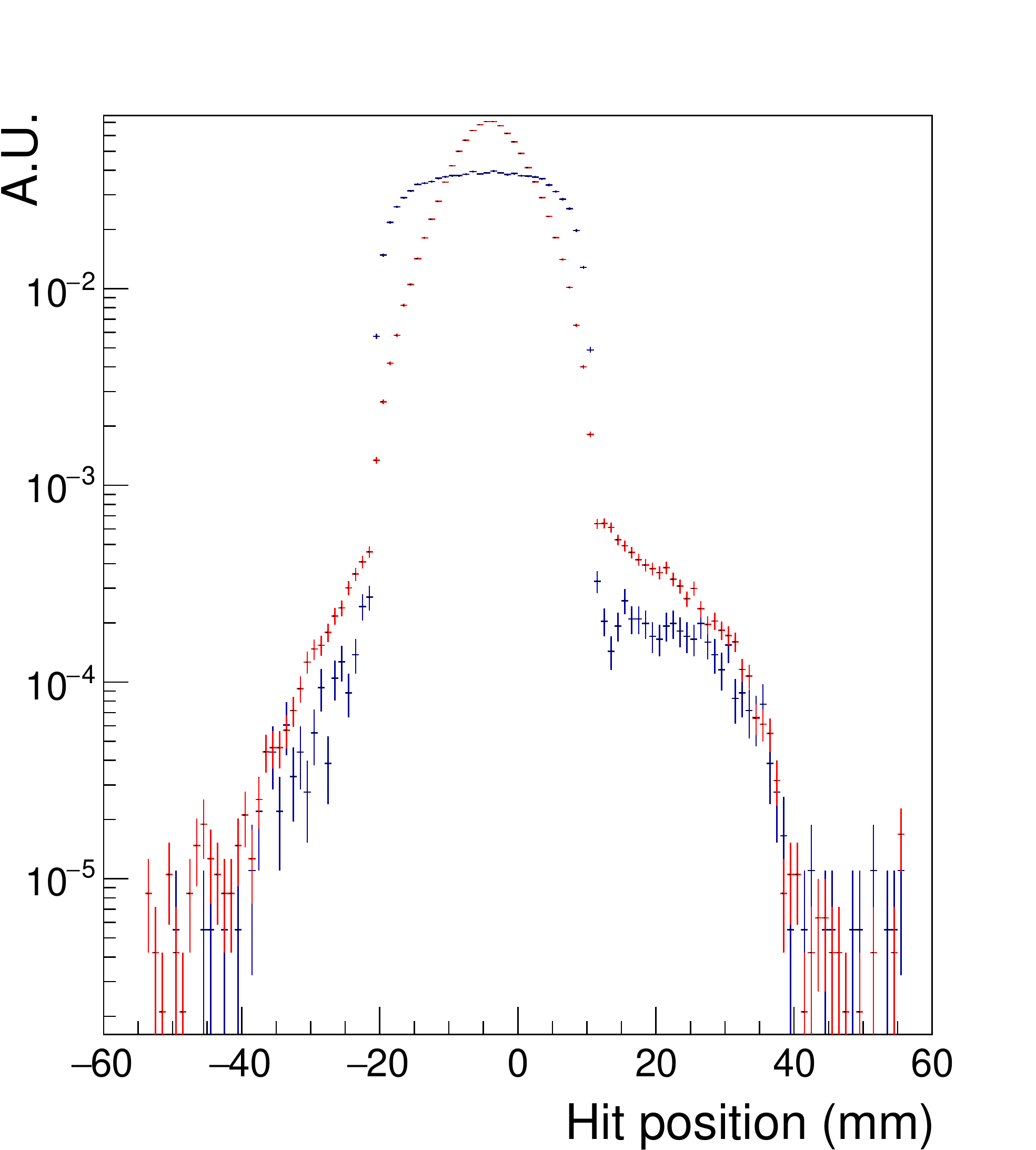}
    \caption{Measured beam profile with the most upstream Micromega detector (MM1) during an electron (red) and a hadron (blue) calibration run. The hadron beam profile width is significantly larger than the electron beam one due to the defocusing effects mentioned in the text. The squared shape of the hadron profile reflects the geometrical acceptance of the scintillators counters in the trigger ($\Phi_{S_0}=3.2$~cm), suggesting that the hadronic beam width is actually even larger than the size of these detectors. }
    \label{fig:6895_6896_MM1Y}
\end{figure}

We also observed that, for the negative-charge mode configuration of the beamline, the $h/e$ ratio differs between the runs. This effect is due to the different beamline configurations used for each run pair, particularly concerning the opening of the beam-defining collimators upstream of the NA64 detector. These slits were adjusted during the data-taking periods to maximize the total beam intensity\footnote{Each configuration corresponds to a dedicated file in the CESAR control system~\cite{Rae:2022szb}.}. The hadrons in the beam are less focused than the electrons due to various effects (synchrotron radiation effects, hadrons interaction along the length of the beamline, pions decay to muons, $\ldots$), as also highlighted by Fig.~\ref{fig:6895_6896_MM1Y}, showing the beam profile measured with the most upstream Micromega detector (MM1) during an electron and a hadron calibration run.
Therefore, the more the collimators are open, the higher number of hadrons are transmitted, and the larger is the $h/e$ contamination. To highlight this effect, in table~\ref{table:runs} we grouped together the data-taking runs referring to the same beamline configuration, showing a compatible value of $h/e$. 

\subsection{Relative fraction of hadronic contaminants}

When the H4 beamline is operated in negative charge mode, the main contribution to the hadronic contamination at 100 GeV/c comes from $\pi^-$ from the $K_S\rightarrow \pi^+\pi^-$ decay, while in positive charge mode it is due to protons from the $\Lambda \rightarrow p\pi^-$ decay. Specifically, at 100 GeV, the differential particle yields per impinging proton $\frac{dN}{dE}$ after the lead converter predicted by FLUKA read $\simeq 2.7\times10^{-7}$~GeV$^{-1}$ (protons), $\simeq 1.2\times10^{-8}$~GeV$^{-1}$ ($\pi^+$), $\simeq 4.2\times10^{-9}$~GeV$^{-1}$ (anti-proton), and $\simeq 1.2\times10^{-8}$~GeV$^{-1}$ ($\pi^-$) (see also Fig.~\ref{fig:T4_sim2}). From this, the predicted fraction of pions with respect to the total number of hadrons \textit{at the lead converter} is $\simeq 74\%$ ($\simeq 4\%$) in negative (positive) mode. In all cases, the yield of kaons is smaller than $10^{-10}$~GeV$^{-1}$.

\begin{table}[t]
    \centering
    \begin{tabular}{c c c}\hline\hline
    H4 config. & Charge & $f_\pi$ (\%)  \\
    \hline
    (a) & Negative & 68 $\pm$ 5 \\  
    (c) & Negative & 85 $\pm$ 4 \\
    (d) & Negative & 80 $\pm$ 5 \\
    (e) & Positive & 30 $\pm$ 4\\
    \end{tabular}
    \caption{Fraction of pions with respect to the total number of hadron contaminants for the different H4 beamline settings. The uncertainty is purely statistical. For the period (b), the statistics of ``punch-throught'' events is too low to allow a proper evaluation of $f_\pi$.}
    \label{tab:hcontamination}
\end{table}

We validated this result by exploiting the different absorption probabilities of protons, anti-protons and pions in matter~\cite{Carroll:1978hc}, and therefore the different absorption probability in the first HCAL module. For illustration, at 100 GeV/c the absorption cross section of these particles in iron, computed from the data in the aforementioned reference, is about 550~mbarn ($\pi$), 690~mbarn (proton) and 720~mbarn (anti-proton), resulting in an absorption length in this material of 21 cm ($\pi$), 17 cm (proton), and 16 cm (anti-proton). We exploited this difference by evaluating, in hadron calibration runs, the fraction of ``punch-trough`` events $f_{punch-through}$ satisfying the $S_{E}$ cut that have a MIP-like signature in the HCAL-1 and with full-energy deposition in HCAL-2, with respect to the total number of events satisfying the $S_{E}$ and the $S_{H}$ cuts. In doing so, we grouped together runs corresponding to the same H4 beamline settings. We also applied a MC-derived correction to $f_{punch-through}$, of about 1.5$\%$, to account for muon-induced events with a ``punch-through'' signature due to Bremmstralung emission in HCAL-2. We compared this result with the punch-through probability $P^{MC}$ from Monte Carlo simulation of the NA64-$e$ setup for each hadron type, $P^{MC}_{p}$ and $P^{MC}_\pi$, where $p$ is either a proton (positive-charge mode run) or an anti-proton (negative-charge mode runs). Finally, we extracted the fraction of pions  among the H4 beamline hadronic contaminants ($f_\pi$) by solving the equation:
\begin{equation}
    f_{punch-through} = f_\pi P^{MC}_{\pi} + f_p P^{MC}_p \; \; ,
\end{equation}
with the constraint $f_\pi+f_p=1$ (i.e., we ignored the residual contributions from kaons). To evaluate the systematic uncertainty associated with the energy thresholds in the $S_{E}$ cut, as before we repeated the evaluation of $f_{punch-through}$ for each combination, and then we computed the standard deviation of all values: the obtained uncertainty is negligible with respect to the statistical one. Similarly, to evaluate the uncertainty associated with the simulation, we repeated the calculation of $f_\pi$ using results obtained from GEANT4 (\texttt{FTFP\_BERT} physics list) and FLUKA; obtained values were compatible within $\simeq 20\%$ ($\simeq 10\%$) for the negative (positive) charge mode.
Our results for $f_{\pi}$ are summarized in Tab.~\ref{tab:hcontamination}; the large values of the statistical uncertainty are due to the very low yield of ``punch-through'' events. The obtained results confirm the trend predicted by Monte Carlo. We ascribed the difference between data and MC in positive-charge mode to the fact that, due to mass-dependent effects such as syncrotron radiation emission, protons are transported by the H4 beamline with lower efficiency than $\pi^+$, thus increasing the measured value of $f_\pi$ at the NA64 detector location.

\section{Conclusions}

We measured the intrinsic hadron contamination of the H4 $e^-/e^+$ beam at CERN. Our analysis exploits data collected by the NA64 experiment during pairs of open-trigger runs with and without the lead converter downstream the T2 target. Comparing these data, we could measure the $h/e$ ratio through a fair methodology, negligibly affected by the systematic uncertainty associated to absolute events normalization. Our experimental results were compared with the prediction from Monte Carlo simulations and a good agreement was found. A further improvement of this prediction would require to introduce in the simulations the effect of the $\sim 540$ m-long beamline between the T2 target and the detector, including non-ideal effects associated to displacements, and goes beyond the scope of this work. 

In conclusion, in this work we proved that the possible electron purities available in the H4 beamline of the SPS North Area may reach, with this tuning, a $e/h$ ratio up to 99.7$\%$ for 100 GeV/c beams, for intensities up to $\simeq 6\times10^6 $ $e^-$/spill. This result is crucial for NA64, as well as any future fixed-target experiment at CERN demanding for high-purity, high-intensity electron beam in this energy range.

\section*{Acknowledgments}
We gratefully acknowledge the support of the CERN management and staff and the technical staffs of the participating institutions for their vital contributions. 
This result is part of a project that has received funding from the European Research Council (ERC) under the European Union's Horizon 2020 research and innovation programme, Grant agreement No. 947715 (POKER). 
This work was supported by the Istituto Nazionale di Fisica Nucleare (Italy), the HISKP, University of Bonn (Germany), ETH Zurich and SNSF Grant No. 169133, 186181, 186158, 197346 (Switzerland), and grants FONDECYT 1191103, 1190845, and 1230160, UTFSM PI~M~18~13 and ANID PIA/APOYO AFB180002, AFB220004 and ANID - Millenium Science Initiative Program - ICN2019\_044 (Chile), and  RyC-030551-I and PID2021-123955NA-100 funded by MCIN/AEI/ 10.13039/501100011033/FEDER, UE (Spain). This work is (artially supported by ICSC – Centro Nazionale di Ricerca in High Performance Computing, Big Data and Quantum Computing, funded by European Union – NextGenerationEU


\bibliography{bibliographyNA64_inspiresFormat,bibliographyNA64exp_inspiresFormat,bibliographyOther_inspiresFormat}

\begin{thebibliography}{10}
\expandafter\ifx\csname url\endcsname\relax
  \def\url#1{\texttt{#1}}\fi
\expandafter\ifx\csname urlprefix\endcsname\relax\def\urlprefix{URL }\fi
\expandafter\ifx\csname href\endcsname\relax
  \def\href#1#2{#2} \def\path#1{#1}\fi

\bibitem{Atherton:164934}
H.~W. Atherton, P.~Coet, N.~T. Doble, D.~E. Plane,
  \href{https://cds.cern.ch/record/164934}{{Electron and photon beams in the
  SPS experimental areas}}, Tech. rep., CERN, Geneva (1985).
\newline\urlprefix\url{https://cds.cern.ch/record/164934}

\bibitem{Banerjee:2774716}
D.~Banerjee, J.~Bernhard, M.~Brugger, N.~Charitonidis, N.~Doble, L.~Gatignon,
  A.~Gerbershagen, \href{https://cds.cern.ch/record/2774716}{{The North
  Experimental Area at the Cern Super Proton Synchrotron}}Dedicated to Giorgio
  Brianti on the 50th anniversary of his founding the SPS Experimental Areas
  Group of CERN-Lab II and hence initiating the present Enterprise. (2021).
\newline\urlprefix\url{https://cds.cern.ch/record/2774716}

\bibitem{Fabbrichesi:2020wbt}
M.~Fabbrichesi, E.~Gabrielli, G.~Lanfranchi, {The Dark Photon}, 2020,
  {SpringerBrief in Physics}.
\newblock \href {http://arxiv.org/abs/2005.01515} {\path{arXiv:2005.01515}},
  \href {https://doi.org/10.1007/978-3-030-62519-1}
  {\path{doi:10.1007/978-3-030-62519-1}}.

\bibitem{Filippi:2020kii}
A.~Filippi, M.~De~Napoli, {Searching in the dark: the hunt for the dark
  photon}, Rev. Phys. 5 (2020) 100042.
\newblock \href {http://arxiv.org/abs/2006.04640} {\path{arXiv:2006.04640}},
  \href {https://doi.org/10.1016/j.revip.2020.100042}
  {\path{doi:10.1016/j.revip.2020.100042}}.

\bibitem{Graham:2021ggy}
M.~Graham, C.~Hearty, M.~Williams, {Searches for dark photons at accelerators},
  2021.
\newblock \href {http://arxiv.org/abs/2104.10280} {\path{arXiv:2104.10280}}.

\bibitem{NA64:2019imj}
D.~Banerjee, et~al., {Dark matter search in missing energy events with NA64},
  Phys. Rev. Lett. 123~(12) (2019) 121801.
\newblock \href {http://arxiv.org/abs/1906.00176} {\path{arXiv:1906.00176}},
  \href {https://doi.org/10.1103/PhysRevLett.123.121801}
  {\path{doi:10.1103/PhysRevLett.123.121801}}.

\bibitem{Volkov:2019qhb}
V.~Volkov, P.~Volkov, T.~Enik, G.~Kekelidze, V.~Kramarenko, V.~Lysan,
  D.~Peshekhonov, A.~Solin, A.~Solin, {Straw Chambers for the NA64 Experiment},
  Phys. Part. Nucl. Lett. 16~(6) (2019) 847--858.
\newblock \href {https://doi.org/10.1134/S1547477119060554}
  {\path{doi:10.1134/S1547477119060554}}.

\bibitem{Depero:2017mrr}
E.~Depero, et~al., {High purity 100 GeV electron identification with
  synchrotron radiation}, Nucl. Instrum. Meth. A 866 (2017) 196--201.
\newblock \href {http://arxiv.org/abs/1703.05993} {\path{arXiv:1703.05993}},
  \href {https://doi.org/10.1016/j.nima.2017.05.028}
  {\path{doi:10.1016/j.nima.2017.05.028}}.

\bibitem{Brianti:604383}
G.~Brianti, \href{https://cds.cern.ch/record/604383}{{SPS North Experimental
  Area}}, Tech. rep., CERN, Geneva (1973).
\newline\urlprefix\url{https://cds.cern.ch/record/604383}

\bibitem{Booth:2019brj}
A.~C. Booth, N.~Charitonidis, P.~Chatzidaki, Y.~Karyotakis, E.~Nowak,
  I.~Ortega-Ruiz, M.~Rosenthal, P.~Sala, {Particle production, transport, and
  identification in the regime of 1\ensuremath{-}7 GeV/c}, Phys. Rev. Accel.
  Beams 22~(6) (2019) 061003.
\newblock \href {https://doi.org/10.1103/PhysRevAccelBeams.22.061003}
  {\path{doi:10.1103/PhysRevAccelBeams.22.061003}}.

\bibitem{Fraser:2019mpj}
M.~Fraser, et~al., {SPS slow extraction losses and activation: update on recent
  improvements}, in: {10th International Particle Accelerator Conference},
  2019, p. WEPMP031.
\newblock \href {https://doi.org/10.18429/JACoW-IPAC2019-WEPMP031}
  {\path{doi:10.18429/JACoW-IPAC2019-WEPMP031}}.

\bibitem{Keizer:319359}
R.~L. Keizer, M.~Mottier, \href{https://cds.cern.ch/record/319359}{{The MTN and
  MTR bending magnets}}, Tech. rep., CERN, Geneva (1978).
\newline\urlprefix\url{https://cds.cern.ch/record/319359}

\bibitem{Ferrari:2005zk}
A.~Ferrari, P.~R. Sala, A.~Fasso, J.~Ranft, {FLUKA: A multi-particle transport
  code (Program version 2005)} (10 2005).
\newblock \href {https://doi.org/10.2172/877507} {\path{doi:10.2172/877507}}.

\bibitem{Bohlen:2014buj}
T.~T. B\"ohlen, F.~Cerutti, M.~P.~W. Chin, A.~Fass\`o, A.~Ferrari, P.~G.
  Ortega, A.~Mairani, P.~R. Sala, G.~Smirnov, V.~Vlachoudis, {The FLUKA Code:
  Developments and Challenges for High Energy and Medical Applications}, Nucl.
  Data Sheets 120 (2014) 211--214.
\newblock \href {https://doi.org/10.1016/j.nds.2014.07.049}
  {\path{doi:10.1016/j.nds.2014.07.049}}.

\bibitem{Agostinelli:2002hh}
S.~Agostinelli, et~al., {GEANT4--a simulation toolkit}, Nucl. Instrum. Meth. A
  506 (2003) 250--303.
\newblock \href {https://doi.org/10.1016/S0168-9002(03)01368-8}
  {\path{doi:10.1016/S0168-9002(03)01368-8}}.

\bibitem{Allison:2006ve}
J.~Allison, et~al., {Geant4 developments and applications}, IEEE Trans. Nucl.
  Sci. 53 (2006) 270.
\newblock \href {https://doi.org/10.1109/TNS.2006.869826}
  {\path{doi:10.1109/TNS.2006.869826}}.

\bibitem{Rae:2022szb}
B.~Rae, et~al., {Controlling the CERN Experimental Area Beams}, in: {18th
  International Conference on Accelerator and Large Experimental Physics
  Control Systems}, 2022.
\newblock \href {http://arxiv.org/abs/2202.01705} {\path{arXiv:2202.01705}}.

\bibitem{Carroll:1978hc}
A.~S. Carroll, et~al., {Absorption Cross-Sections of $\pi^{\pm}$, $K^{\pm}$, p
  and $\bar{p}$ on Nuclei Between 60 GeV/c and 280 GeV/c}, Phys. Lett. B 80
  (1979) 319--322.
\newblock \href {https://doi.org/10.1016/0370-2693(79)90226-0}
  {\path{doi:10.1016/0370-2693(79)90226-0}}.

\end{thebibliography}

\end{document}